\documentclass[sigconf]{acmart}

\usepackage{booktabs} % For formal tables,
\usepackage{xcite}
\usepackage{xr}
\usepackage{graphicx}
\usepackage{caption}
\usepackage{subcaption}
\usepackage{multirow}
\usepackage{bigstrut}
\usepackage{enumitem}
\usepackage[skip=0pt]{caption}
\usepackage[ruled,linesnumbered]{algorithm2e}
\usepackage{bm}
\usepackage{tabularx,ragged2e}
\usepackage{xcolor}
\usepackage{verbatim}

\externalcitedocument{introduction}
\externalcitedocument{methods}
\externalcitedocument{evaluation}
\renewcommand\footnotetextcopyrightpermission[1]{} % removes footnote with conference information in first column
\makeatletter
\def\subsubsection{\@startsection{subsubsection}{3}%
  \z@{.3\linespacing\@plus.7\linespacing}{.1\linespacing}%
  {\normalfont\itshape}}
\makeatother

\copyrightyear{2019} 
\acmYear{2019} 
\setcopyright{acmcopyright}
\acmConference[WSDM '19]{The Twelfth ACM International Conference on Web Search and Data Mining}{February 11--15, 2019}{Melbourne, VIC, Australia}
\acmBooktitle{The Twelfth ACM International Conference on Web Search and Data Mining (WSDM '19), February 11--15, 2019, Melbourne, VIC, Australia}
\acmPrice{15.00}
\acmDOI{10.1145/3289600.3290977}
\acmISBN{978-1-4503-5940-5/19/02}

\begin{document}
\title{Gated Attentive-Autoencoder for Content-Aware Recommendation}

\author{Chen Ma}
\affiliation{\institution{McGill University}}
\email{chen.ma2@mail.mcgill.ca}

\author{Peng Kang}
\affiliation{\institution{McGill University}}
\email{peng.kang2@mail.mcgill.ca}

\author{Bin Wu}
\affiliation{\institution{Zhengzhou University}}
\email{wubin@gs.zzu.edu.cn}

\author{Qinglong Wang}
\affiliation{\institution{McGill University}}
\email{qinglong.wang@mail.mcgill.ca}

\author{Xue Liu}
\affiliation{\institution{McGill University}}
\email{xueliu@cs.mcgill.ca}

\begin{abstract}
The rapid growth of Internet services and mobile devices provides an excellent opportunity to satisfy the strong demand for the personalized item or product recommendation. However, with the tremendous increase of users and items, personalized recommender systems still face several challenging problems: (1) the hardness of exploiting sparse implicit feedback; (2) the difficulty of combining heterogeneous data. To cope with these challenges, we propose a gated attentive-autoencoder (GATE) model, which is capable of learning fused hidden representations of items' contents and binary ratings, through a neural gating structure. Based on the fused representations, our model exploits neighboring relations between items to help infer users' preferences. In particular, a word-level and a neighbor-level attention module are integrated with the autoencoder. The word-level attention learns the item hidden representations from items' word sequences, while favoring informative words by assigning larger attention weights. The neighbor-level attention learns the hidden representation of an item's neighborhood by considering its neighbors in a weighted manner. We extensively evaluate our model with several state-of-the-art methods and different validation metrics on four real-world datasets. The experimental results not only demonstrate the effectiveness of our model on top-N recommendation but also provide interpretable results attributed to the attention modules.
\end{abstract}

% We no longer use \terms command
%\terms{Theory}

% \keywords{Content-aware Recommendation; Autoencoders; Attention Mechanism}

\maketitle

\section{Introduction}
With the rapid growth of Internet services and mobile devices, it has been more convenient for people to access amounts of online products and multimedia contents, such as movies and articles. Although this growth allows users to have multiple choices, it has also made it more difficult to select one of the user's most preferred items out of thousands of candidates. For example, users who like to watch movies may feel difficult to decide which movie to watch when there are thousands of selections, and users who are gourmet eaters may feel hard to discover new restaurants tailored to their flavors. Therefore, these needs facilitate a promising service--personalized recommender systems. These systems are becoming increasingly essential, serving a potentially huge service demand, and bringing significant benefits to at least two parties: (1) help users easily discover products that they are interested in; (2) create opportunities for product providers to increase the revenue.

To build personalized recommender systems, two types of data are generally available and utilized: user ratings and item descriptions, e.g., users' ratings on movies and movies' plots. Approaches based on item text modeling such as latent dirichlet allocation (LDA), stacked denoising autoencoder (SDAE), and variational autoencoder (VAE) have been proposed to additionally utilize items' descriptions, e.g., reviews, abstracts, or synopses \cite{DBLP:conf/kdd/WangB11,DBLP:conf/kdd/WangWY15,DBLP:conf/kdd/LiS17}, to enhance the top-N recommendation performance. Collaborative deep learning (CDL) \cite{DBLP:conf/kdd/WangWY15} and collaborative variational autoencoder (CVAE) \cite{DBLP:conf/kdd/LiS17} are two representative methods, which explicitly link the learning of item content to the recommendation task. In particular, CVAE and CDL apply a VAE and an SDAE, respectively, to learn hidden representations from items' bag-of-words, which are integrated with the probabilistic matrix factorization (PMF) by regularizing with PMF's item latent factors. 

Although existing methods have proposed effective models and achieved satisfactory results, we argue that there are still several factors to be considered for enhancing the performance. First, previous studies \cite{DBLP:conf/kdd/WangWY15,DBLP:conf/kdd/LiS17} learn the content hidden representations from items' normalized bag-of-words vectors, which does not consider the importances of different words for describing a certain item. Equally treating the informative words along with other words may lead to the incomplete understanding of the item content. Second, previous works \cite{DBLP:conf/recsys/KimPOLY16,DBLP:conf/kdd/LiS17} combine the hidden representations from heterogeneous information, e.g., items' ratings and descriptions, by a weighted regularization term. This may not fully make use of the data from heterogeneous sources and trigger tedious hyper-parameter tuning,  since different data sources are characterized by distinct statistical properties and different orders of magnitude, which is commonly the case for heterogeneous information. Third, it is also important to note that the relations between items, e.g., movies in the same genre and citations between articles, are neglected in previous works. It is very likely that closely related items may share the same topics or have similar attributes. As such, exploring users' preferences on an item's neighbors also benefits inferring users' preferences on this item.

To address the problems mentioned above, we propose a novel recommendation model, gated attentive-autoencoder (GATE), for the content-aware recommendation. GATE consists of a word-attention module, a neighbor-attention module, and a neural gating structure, integrating with a stacked autoencoder (AE). The encoder of the stacked AE encodes the user's implicit feedback on a certain item into the item's hidden representation. Then the word-attention module learns the item embedding from its sequence of words, where the informative words can be adaptively selected without using complex recurrent or convolutional neural networks. To smoothly fuse the representations of items' ratings and descriptions, we propose a neural gating layer to extract and merge the salient parts of these two hidden representations, which is inspired by the long short-term memory (LSTM) \cite{DBLP:journals/neco/HochreiterS97}. Moreover, item-item relations provide important auxiliary information to predict users' preferences, since closely related items may have the same topics or attributes. Thus, we apply a neighbor-attention module to learn the hidden representation of an item's neighborhood. By modeling users' preferences on the item's neighborhood, the users' preferences on this item can be indirectly reflected. We extensively evaluate our model with many state-of-the-art methods and different validation metrics on four real-world datasets. The experimental results not only demonstrate the improvements of our model over other baselines but also show the effectiveness of the gating layer and attention modules. 

To summarize, the major contributions of this paper are listed as follows:
\begin{itemize}[leftmargin=*]
\item To learn the hidden representations from items' sequences of words, we apply a word-attention module to adaptively distinguish informative words, leading to better comprehension of the item content. Our word-attention module can achieve the same performance with complex recurrent or convolutional neural networks yet with fewer parameters.
\item To effectively fuse the hidden representations of items' contents and ratings, we propose a neural gating layer to extract and combine the salient parts of them. %We validate that the gating layer achieves better results than combining these two representations by regularization.
\item According to item-item relations, we utilize a neighbor-attention module to learn the hidden representation of an item's neighborhood. Modeling user preferences on the neighborhood of an item provides a significant supplement for inferring user preferences on this item.
\item Both proposed attention modules are capable of interpreting and visualizing the important words and neighbors of items, respectively. Experiments on four real-world datasets show that the proposed GATE model significantly outperforms the state-of-the-art methods for the content-aware recommendation.
\end{itemize}

\section{Related Work}
In this section, we illustrate related work about the proposed model: personalized recommendation, recommendation with content features, and attention mechanisms.

\subsection{Recommendation with Implicit Feedback}
Early studies on recommendation have largely focused on explicit feedback \cite{DBLP:conf/www/SarwarKKR01,DBLP:conf/icml/SalakhutdinovMH07}, recent research focus is much shifting towards implicit data \cite{DBLP:conf/kdd/WangWY15,DBLP:conf/kdd/LiS17}. The collaborative filtering (CF) with implicit feedback is usually treated as a top-N item recommendation task, where the goal is to recommend a list of items to users that users may be interested in. Compared to the rating prediction task, the item recommendation problem is more practical and challenging \cite{DBLP:conf/icdm/PanZCLLSY08}, which more accords with the real-world recommendation scenario. To make use of latent factor models for item recommendation, early works either applied a uniform weighting scheme to treat all missing data as negative samples \cite{DBLP:conf/icdm/HuKV08}, or sampled negative instances from missing data \cite{DBLP:conf/uai/RendleFGS09}. Recently, He et al. \cite{DBLP:conf/sigir/HeZKC16} and Liang et al. \cite{DBLP:conf/www/LiangCMB16} proposed dedicated models to weigh missing data, and Bayer et al. \cite{DBLP:conf/www/BayerHKR17} developed an implicit coordinate descent method for feature-based factorization models. With the ability to learn salient representations, (deep) neural network-based methods were also adopted. In \cite{DBLP:conf/wsdm/WuDZE16}, Wu et al. proposed the collaborative denoising autoencoder (CDAE) for top-N recommendation learning from implicit feedback. In \cite{DBLP:conf/www/HeLZNHC17}, He et al. proposed a neural network-based collaborative filtering model, which leverages a multi-layer perceptron to learn the non-linear user-item interactions. In \cite{DBLP:conf/ijcai/XueDZHC17,DBLP:conf/ijcai/GuoTYLH17,DBLP:conf/kdd/LianZZCXS18}, deep learning techniques were adopted to boost the traditional matrix factorization and factorization machine methods.

\subsection{Content-Aware Recommendation}
To further improve the performance, researchers incorporate the content features to help alleviate the sparseness and the cold-start problem in the user-item interaction data. Some early works \cite{DBLP:conf/recsys/McAuleyL13,DBLP:conf/kdd/WangB11} applied latent dirichlet allocation (LDA) to learn abstract topics that occur in a collection of documents. In recent years, deep learning models have demonstrated a great power for effective text representation learning. In \cite{DBLP:conf/kdd/WangWY15,DBLP:conf/kdd/ZhangYLXM16}, researchers utilized the stacked denoising autoencoder (SDAE) on items' bag-of-words to learn the item latent representations. Li et al. in \cite{DBLP:conf/cikm/LiKF15} proposed to combine probabilistic matrix factorization with marginalized denoising autoencoders. And in \cite{DBLP:conf/kdd/LiS17}, Li et al. adopted a variational autoencoder to learn the latent representations from items' content, which is a Bayesian probabilistic generative model. These studies model the item content through its bag-of-words. On the other hand, some studies also incorporate the contextual information for better understanding of the text. For example, in \cite{DBLP:conf/cikm/ZhangACC17}, the doc2vec model \cite{DBLP:conf/icml/LeM14} was utilized to model the text information in user reviews. And in \cite{DBLP:conf/recsys/KimPOLY16,DBLP:conf/recsys/SeoHYL17,DBLP:conf/wsdm/ZhengNY17,DBLP:conf/www/ChenZLM18}, researchers adopted convolution neural networks, with max-pooling and fully connected layers, to learn the item hidden representation from item's sequence of word embeddings, where words' contextual information can be captured by the convolutional filters and the sliding window strategy. 

\subsection{Attention Mechanism in Recommendation}
Recently, attention mechanism has demonstrated the effectiveness in various machine learning tasks such as image captioning \cite{DBLP:conf/icml/XuBKCCSZB15,DBLP:conf/cvpr/YouJWFL16}, document classification \cite{DBLP:conf/naacl/YangYDHSH16}, and machine translation \cite{DBLP:conf/emnlp/LuongPM15,DBLP:conf/nips/VaswaniSPUJGKP17}. Researchers also adopt the attention mechanism on recommendation tasks. In \cite{DBLP:conf/cikm/PeiYSZBT17}, Pei et al. adopted an attention model to measure the relevance between users and items. Wang et al. \cite{DBLP:conf/kdd/WangYRTZYW17} proposed a hybrid attention model to adaptively capture the change of editors' selection criteria. In \cite{DBLP:conf/ijcai/GongZ16}, Gong et al. adopted an attention model to scan input microblogs and select trigger words. Chen et al. \cite{DBLP:conf/sigir/ChenZ0NLC17} proposed item- and component-level attention mechanisms to model the implicit feedback in the multimedia recommendation. In \cite{DBLP:conf/recsys/SeoHYL17}, Seo et al. proposed to model user preferences and item properties using convolutional neural networks (CNNs) with dual local and global attention. In \cite{DBLP:conf/kdd/TayLH18}, Tay et al. proposed a multi-pointer attention mechanism to enhance the rating prediction accuracy. In \cite{DBLP:conf/cikm/MaZWL18}, Ma et al. integrated the attention mechanism with autoencoders to discriminate the user preferences on users' visited locations. And in \cite{DBLP:conf/www/ChenZLM18}, an attention-based review pooling mechanism was proposed to select the important user reviews. 

However, our word- and neighbor-attention modules are different from above studies. For the word-attention module, we adopt the multi-dimensional attention to select informative words by computing a score vector. While the vanilla attention computes a single importance score for each word, which cannot sufficiently express the complex relations among words when the number of words is large. For the neighbor-attention module, we learn item's neighborhood representation according to the importance scores between the item and its neighbors. The item-item relation information is rarely considered in previous works. Moreover, we propose a neural gating layer to adaptively merge items' hidden representations from different data sources.

\section{Problem Formulation}
The recommendation task considered in this paper takes implicit feedback \cite{DBLP:conf/icdm/HuKV08} as the training and test data. The user preferences are presented by an $ m $-by-$ n $ binary matrix $ \mathbf{R} $. The entire collection of $ n $ items is represented by a list of documents $ \mathcal{D} $, where each document in $ \mathcal{D} $ is represented by a sequence of words. The item relations are presented by a binary adjacent matrix $ \mathbf{N} \in \mathbb{R}^{n \times n} $, where $ N_{ij}=1 $ if item $ i $ and $ j $ are related or connected. Given the item descriptions $ \mathcal{D} $, the item relations $ \mathbf{N} $, and part of the ratings in $ \mathbf{R} $, the problem is to predict the rest of ratings in $ \mathbf{R} $.

Here, following common symbolic notation, upper case bold letters denote matrices, lower case bold letters denote column vectors without any specification, and non-bold letters represent scalars. 

\begin{figure*}[t!]
    \centering
    \includegraphics[width=0.9\linewidth]{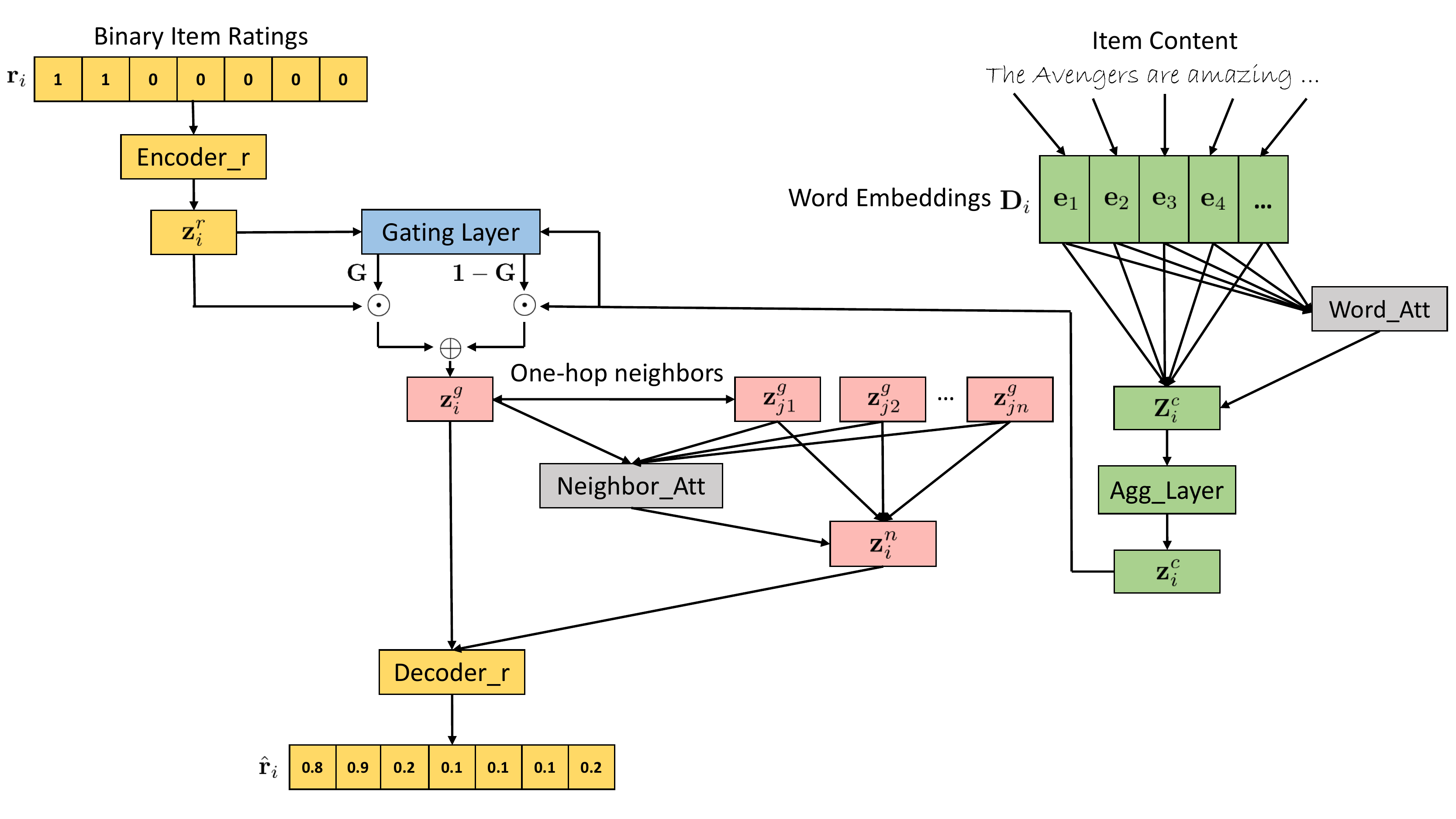}
    \caption{\label{fig:whole_model}The architecture of GATE. The yellow part is the stacked AE for binary rating prediction, and the green part is the word-attention module for item content. The blue rectangle is the gating layer to fuse the hidden representations. The middle pink part is the neighbor-attention module to obtain the hidden representation of an item's neighborhood. Specifically, \textit{Word\_Att} denotes the word-attention layer, \textit{Neighbor\_Att} denotes the neighbor-attention layer, and \textit{Agg\_Layer} denotes the aggregation layer. $ \odot $ is the element-wise multiplication and $ \oplus $ is the element-wise addition.}
\vspace{-0.3cm}
\end{figure*}

\section{Methodologies}
In this section, we introduce the proposed model, which is shown in Figure \ref{fig:whole_model}. We first illustrate the basic model to learn item representations from users' binary ratings. We then introduce the multi-dimensional attention for learning item representations from word sequences. Next, we present the neural gating layer to combine the item representations from ratings and contents. We then demonstrate how to learn the hidden representation of an item's neighborhood and utilize it to assist in inferring user preferences. Lastly, we go through the loss function and training process of the proposed model.

\subsection{Model Basics}
The substantial increase of users and items makes the user-item interactions more complex and hard to model. Classical matrix factorization (MF) methods apply the inner product to predict user preferences on items, which linearly combines users' and items' latent factors. However, it has been shown in \cite{DBLP:conf/www/HeLZNHC17,DBLP:conf/www/HsiehYCLBE17} how the linear combination of the inner product can limit the expressiveness of MF. Inspired by the recent works using autoencoders (AEs) to model explicit feedback \cite{DBLP:conf/www/SedhainMSX15} and implicit feedback \cite{DBLP:conf/wsdm/WuDZE16}, we also adopt AE as our base building block due to its ability to learn richer representations and the close relationship to MF \cite{DBLP:conf/wsdm/WuDZE16}.

To capture users' preferences on an item, we apply a stacked AE to encode users' binary ratings $ \mathbf{r}_{i} \in \mathbb{R}^{m} $ on a certain item $ i $ into the item's rating hidden representation $ \mathbf{z}_{i}^{r} $ (the superscript $ r $ indicates the hidden representation is learned from items' binary ratings):
\begin{equation}
enc:
\begin{cases}
\mathbf{z}_{i}^{(1)} = a_{1}(\mathbf{W}_{1} \mathbf{r}_{i} + \mathbf{b}_{1}) \\
\mathbf{z}_{i}^{r} = a_{2}(\mathbf{W}_{2} \mathbf{z}_{i}^{(1)} + \mathbf{b}_{2})
\end{cases}
dec:
\begin{cases}
\mathbf{z}_{i}^{(3)} = a_{3}(\mathbf{W}_{3} \mathbf{z}_{i}^{r} + \mathbf{b}_{3}) \\
\hat{\mathbf{r}}_{i} = a_{4}(\mathbf{W}_{4} \mathbf{z}_{i}^{(3)} + \mathbf{b}_{4})
\end{cases}
\label{eq:rating_AE}
\end{equation}
where $ \mathbf{W}_{1} \in \mathbb{R}^{h_{1} \times m} $, $ \mathbf{W}_{2} \in \mathbb{R}^{h \times h_{1}} $, $ \mathbf{W}_{3} \in \mathbb{R}^{h_{1} \times h} $, and $ \mathbf{W}_{4} \in \mathbb{R}^{m \times h_{1}} $ are the weight matrices. $ m $ is the number of users, $ h_{1} $ is the dimension of the first hidden layer, and $ h $ is the dimension of the bottleneck layer. $ \mathbf{r}_{i} $ is a multi-hot vector, where $ r_{u,i} = 1 $ indicates that the user $ u $ prefers the item $ i $.

\subsection{Word-Attention Module} \label{sec:word_attention}
Unlike previous works \cite{DBLP:conf/kdd/WangWY15,DBLP:conf/kdd/LiS17,DBLP:conf/www/HsiehYCLBE17} learning item embeddings from bag-of-words and neglecting the importances of different words, we propose a word-attention module based on items' word sequences. Compared to learning from items' bag-of-words, the attention weights learned by our module adaptively select the informative words with different importances, and make the informative words contribute more to depict items.

\textbf{Embedding Layer}. In the proposed module, the input of item $ i $ is a sequence of $ l_i $ words from its text description, where each word is represented as an one-hot vector. At the embedding layer, the one-hot encoded vector is converted into a low-dimensional real-valued dense vector representation by a word embedding matrix $ \mathbf{E} \in \mathbb{R}^{h \times v} $, where $ h $ is the dimension of the word embedding and $ v $ is size of the vocabulary. After converted by the embedding layer, the item text is represented as:
\[ 
\mathbf{D}_i = \begin{bmatrix}
 & | & | & | & \\ 
... & \mathbf{e}_{j-1} & \mathbf{e}_{j} & \mathbf{e}_{j+1} & ... \\ 
 & | & | & | & 
\end{bmatrix}
\]
where $ \mathbf{D}_i \in \mathbb{R}^{h \times l_i} $ and $ \mathbf{e}_{j} \in \mathbb{R}^{h} $.

\textbf{Multi-dimensional Attention}. Inspired by the Transformer \cite{DBLP:conf/nips/VaswaniSPUJGKP17} solely relying on attention mechanisms for machine translation, we apply a multi-dimensional attention mechanism on word sequences to learn items' hidden representations without using complex recurrent or convolutional neural networks. The reason is that, in the real-world scenario, users may care more about the topics or motifs of items that can be illustrated in a few of words, rather than the word-word relations in the sequence.
% By doing this, the same word with distinct surrounding words will be assigned with different attention scores, therefore, contributing to the item hidden representation differently according to the attention scores.

The goal of the word-attention is to assign different importances on words, then aggregate word embeddings in a weighted manner to characterize the item. Given word embeddings of an item $ \mathbf{D}_i $, a vanilla attention mechanism to compute the attention weights is represented by a two-layer neural network:
\begin{equation}
\mathbf{a}_{i} = softmax(\mathbf{w}_{a_1}^{\top} tanh(\mathbf{W}_{a_2} \mathbf{D}_i + \mathbf{b}_{a_2})),
\label{eq:attention_vector}
\end{equation}
where $ \mathbf{w}_{a_1} \in \mathbb{R}^{h} $, $ \mathbf{W}_{a_2} \in \mathbb{R}^{h \times h} $, and $ \mathbf{b}_{a_2} \in \mathbb{R}^{h} $ are the parameters to be learned, the $ softmax(\cdot) $ ensures all the computed weights sum up to 1. Then we sum up the embeddings in $ \mathbf{D}_i $ according to the weights provided by $ \mathbf{a}_{i} $ to get the vector representation of the item (the superscript $ c $ indicates the hidden representation is learned from items' contents):
\begin{equation}
\mathbf{z}_{i}^{c} = \sum_{\mathbf{e}_j \in D_i} a_{i,j} \mathbf{e}_j.
\end{equation}

However, assigning a single importance value to a word embedding usually makes the model focus on a specific aspect of an item content \cite{DBLP:journals/corr/LinFSYXZB17}. It can be multiple aspects in the item content that together characterize this item, especially when the number of words is large. % For example, by adopting a single importance score, the attention mechanism may not easily distinguish the meanings of the same word in different contexts \cite{DBLP:conf/aaai/ShenZLJPZ18}.

Thus, we need multiple $ \mathbf{a}_{i} $ to focus on different parts of the item content. Based on this inspiration, we adopt a matrix instead of $ \mathbf{a}_{i} $ to capture the multi-dimensional attention and assign an attention weight vector to each word embedding. Each dimension of the attention weight vector represents an aspect of relations among all embeddings in $ \mathbf{D}_i $. Suppose we want $ d_{a} $ aspects of attention to be extracted from the embeddings, then we extend $ \mathbf{w}_{a} $ to $ \mathbf{W}_{a_1} \in \mathbb{R}^{d_{a} \times h} $, which behaves like a high level representation of a fixed query "what are the informative words" over other words in the text:
\begin{equation}
\mathbf{A}_{i} = softmax(\mathbf{W}_{a_1} tanh(\mathbf{W}_{a_2} \mathbf{D}_i + \mathbf{b}_{a_2}) + \mathbf{b}_{a_1}),
\label{eq:attention_matrix}
\end{equation}
where $ \mathbf{A}_{i} \in \mathbb{R}^{d_{a} \times l_i} $ is the attention weight matrix, $ \mathbf{b}_{a_1} \in \mathbb{R}^{d_a} $ is the bias term, and the $ softmax $ is performed along the second dimension of its input. By multiplying the attention weight matrix with word embeddings, we have the matrix representation of an item:
\begin{equation}
\mathbf{Z}_{i}^{c} = \mathbf{A}_{i} \mathbf{D}^{\top}_i,
\label{eq:item_embedding_matrix}
\end{equation}
where $ \mathbf{Z}_{i}^{c} \in \mathbb{R}^{d_{a} \times h} $ is the matrix representation of the item. Then we have another neural layer to aggregate the item matrix representation into a vector representation. The hidden representation of the item is revised as:
\begin{equation}
\mathbf{z}_{i}^{c} = a_{t}(\mathbf{Z}_{i}^{c \top} \mathbf{w}_{t}),
\label{eq:item_content_embedding}
\end{equation}
where $ \mathbf{w}_{t} \in \mathbb{R}^{d_{a}} $ is the parameter in the aggregation layer, $ a_{t}(\cdot) $ is the activation function.

\subsection{Neural Gating Layer}
We have obtained the item hidden representations from two heterogeneous data sources, i.e., the binary ratings and the content descriptions of items. The next aim is to combine these two kinds of hidden representations to facilitate the user preference prediction on unrated items. Unlike previous works \cite{DBLP:conf/kdd/WangWY15,DBLP:conf/kdd/LiS17} regularizing these two kinds of hidden representations, we propose a neural gating layer to adaptively merge them. This is inspired by the gates in long short-term memory (LSTM) \cite{DBLP:journals/neco/HochreiterS97}. The gate $ \mathbf{G} $ and the fused item hidden representation $ \mathbf{z}_{i}^{g} $ are computed by:
\begin{equation}
\begin{aligned}
& \mathbf{G} = sigmoid(\mathbf{W}_{g_1} \mathbf{z}_{i}^{r} + \mathbf{W}_{g_2} \mathbf{z}_{i}^{c} + \mathbf{b}_{g}), \\
& \mathbf{z}_{i}^{g} = \mathbf{G} \odot \mathbf{z}_{i}^{r} + (\mathbf{1} - \mathbf{G}) \odot \mathbf{z}_{i}^{c},
\end{aligned}
\label{eq:gated_fusion}
\end{equation}
where $ \mathbf{W}_{g_1} \in \mathbb{R}^{h \times h} $, $ \mathbf{W}_{g_2} \in \mathbb{R}^{h \times h} $, and $ \mathbf{b}_{g} \in \mathbb{R}^{h} $ are the parameters in the gating layer. By using a gating layer, the salient parts from these two hidden representations can be extracted and smoothly combined.

\subsection{Neighbor-Attention Module}
Some items have the inherent relationship between each other, e.g., paper citations. Those closely related items may form a local neighborhood that shares the same topic or has the same attributes. Therefore, for a certain item, if a user is interested in its neighborhood, the user may also be interested in this item. Besides, in an item's local neighborhood, some items may be more representative, which should play an important role in describing the neighborhood. Inspired by this intuition, we propose a neighbor-attention module to learn the neighborhood hidden representation of a certain item. This attention mechanism is similar to which in the machine translation \cite{DBLP:conf/emnlp/LuongPM15}.

Formally, we define the neighbor set of item $ i $ as $ \mathcal{N}_{i} $, which can be obtained from the item adjacent matrix\footnote{For items that do not inherently have item-item relations, we can compute the item-item similarity from the binary rating matrix $ \mathbf{R} $ and set a threshold to select neighbors.} $ \mathbf{N} $. The neighborhood hidden representation $ \mathbf{z}_{i}^{n} $ of item $ i $ is computed by:
\begin{equation}
\begin{aligned}
& s_{i,j} = tanh(\mathbf{z}_{i}^{g \top} \mathbf{W}_{n} \mathbf{z}_{j}^{g}), \forall j \in \mathcal{N}_{i}, \\
& \mathbf{a}_i = softmax(\mathbf{s}_{i}), \\
& \mathbf{z}_{i}^{n} = \sum_{j \in \mathcal{N}_{i}} a_{i,j} \mathbf{z}_{j}^{g}, 
\label{eq:neighbor_attention}
\end{aligned}
\end{equation}
where $ \mathbf{W}_{n} \in \mathbb{R}^{h \times h} $ is the parameters to be learned in the neighbor-attention layer.

To simultaneously capture users' preferences on a certain item and its neighborhood, the decoder in Eq. \ref{eq:rating_AE} is rewritten as:
\begin{equation}
\begin{aligned}
& \mathbf{z}_{i}^{(3,g)} = a_{3}(\mathbf{W}_{3} \mathbf{z}_{i}^{g} + \mathbf{b}_{3}), \\
& \mathbf{z}_{i}^{(3,n)} = a_{3}(\mathbf{W}_{3} \mathbf{z}_{i}^{n} + \mathbf{b}_{3}), \\
& \hat{\mathbf{r}}_{i} = a_{4}(\mathbf{W}_{4} \mathbf{z}_{i}^{(3,g)} + \mathbf{W}_{4} \mathbf{z}_{i}^{(3,n)} + \mathbf{b}_{4}).
\end{aligned}
\label{eq:rating_decoder}
\end{equation}

\subsection{Weighted Loss}
To model the user preference from implicit feedback, we follow a similar manner in \cite{DBLP:conf/icdm/HuKV08} to plug in a confidence matrix in the square loss function:
\begin{equation}
\mathcal{L}_{AE} = \sum_{i=1}^{n} \sum_{u = 1}^{m} ||C_{u,i} (R_{u,i} - \hat{R}_{u,i})||_{2}^{2} = ||\mathbf{C}^{\top} \odot (\mathbf{R}^{\top} - \mathbf{\hat{R}}^{\top})||^{2}_{F},
\label{eq:AE_loss_rating}
\end{equation}
where $ \odot $ is the element-wise multiplication of matrices. $ ||\cdot||_{F} $ is the Frobenius norm of matrices. In particular, we set the confidence matrix $ \mathbf{C} \in \mathbb{R}^{m \times n} $ as follows, 
\begin{equation}
C_{u,i} =
\begin{cases}
\rho & \text{ if } R_{u,i} = 1 \\ 
1 & \text{} otherwise
\end{cases}
\label{eq:rating_weight_function}
\end{equation}
where the hyper-parameter $ \rho > 1 $ is a constant.

\subsection{Network Training}
By combining with regularization terms, the objective function of the proposed model is shown as follows:
\begin{equation}
\begin{aligned}
\mathcal{L} = \mathcal{L}_{AE} + \lambda(||\mathbf{W}_{*}||^{2}_{F} + ||\mathbf{w}_{t}||^{2}_{2}),
\end{aligned}
\label{eq:final_loss}
\end{equation}
where $ \lambda $ is the regularization parameter. By minimizing the objective function, the partial derivatives with respect to all the parameters can be computed by gradient descent with back-propagation. We apply Adam \cite{DBLP:journals/corr/KingmaB14} to automatically adapt the learning rate during the learning procedure. % The mini-batch training algorithm is presented in Alg \ref{alg:batch_training_algorithm}.

\section{Experiments}
In this section, we evaluate the proposed model with the state-of-the-art methods on four real-world datasets.

\subsection{Datasets}
The proposed models are evaluated on four real-world datasets from various domains with different sparsities: \textit{citeulike-a} \cite{DBLP:conf/kdd/WangB11}, \textit{movielens-20M} \cite{DBLP:journals/tiis/HarperK16}, \textit{Amazon-Books} and \textit{Amazon-CDs} \cite{DBLP:conf/www/HeM16}. The \textit{citeulike-a} dataset provides user preferences on articles as well as article titles, abstracts, and citations. The \textit{movielens-20M} is a user-movie dataset where the movie description is crawled from TMDB\footnote{https://www.themoviedb.org/}. The \textit{Amazon-Books} and \textit{Amazon-CDs} datasets are adopted from the Amazon review dataset\footnote{http://jmcauley.ucsd.edu/data/amazon/}, which covers a large amount of user-item interaction data, e.g., review, rating, helpfulness rating of review. We select the user review with the highest helpfulness rating as the item's description. In order to be consistent with the implicit feedback setting, we keep those with ratings no less than four (out of five) as positive feedback and treat all other ratings as missing entries on last three datasets. Since items in the latter three datasets do not inherently have the item-item relations, we compute the item-item similarity from binary rating matrix $ \mathbf{R} $ and set the threshold as 0.2 to select neighbors of items. To filter noisy data, we only keep the users with at least ten ratings and the items at least with five ratings. The data statistics after preprocessing are shown in Table \ref{tab:data_statistics}. For each user, we randomly select 20\% of her rated items as ground truth for testing. The remaining constitutes the training set. The random selection is carried out five times independently, and we report the average results.

\begin{table}[ht]
\centering
\caption{\label{tab:data_statistics}The statistics of datasets.}
\begin{tabular}{ |c|c|c|c|c|c| }
 \hline
 Dataset & \#Users & \#Items & \#Ratings & \#Words & Density \\
 \hline
 \textit{citeulike-a} & 5,551 & 16,980 & 204,986 & 8,000 & 0.217\% \\ 
 \hline
 \textit{ML20M} & 138,493 & 18,307 & 19,977,049 & 12,397 & 0.788\% \\ 
 \hline
 \textit{Books} & 65,476 & 41,264 & 1,947,765 & 27,584 & 0.072\% \\ 
 \hline
 \textit{CDs} & 24,934 & 24,634 & 478,048 & 24,341 & 0.078\% \\ 
 \hline
\end{tabular}
\vspace{-0.3cm}
\end{table}

\subsection{Evaluation Metrics}
We evaluate our model versus other methods in terms of \textit{Recall@k} and \textit{NDCG@k}. For each user, Recall@k (R@k) indicates what percentage of her rated items can emerge in the top $ k $ recommended items. NDCG@k (N@k) is the normalized discounted cumulative gain at $ k $, which takes the position of correctly recommended items into account. 

% comment start
\begin{comment}

\begin{equation}
\begin{aligned}
& P@k =\frac{1}{M} \sum_{i=1}^{M} \frac{S_{i}(k) \cap T_{i}}{k},
R@k =\frac{1}{M} \sum_{i=1}^{M} \frac{S_{i}(k) \cap T_{i}}{|T_{i}|}, \\
& MAP@k =\frac{1}{M} \sum^{M}_{i=1} \frac{\sum^{k}_{j=1} p(j) \times rel(j)}{|T_{i}|},
\end{aligned}
\end{equation}
where $ M $ is number of users, $ S_{i}(k) $ is a set of top-$ k $ unrated items recommended to user $ i $ excluding those items in the training, and $ T_{i} $ is a set of items that are rated by user $ i $ in the testing. $ p(j) $ is the precision of a cut-off rank list from $ 1 $ to $ j $, and $ rel(j) $ is an indicator function that equals to $ 1 $ if the item is rated in the testing, otherwise equals to $ 0 $.

\end{comment}
% comment end

\subsection{Methods Studied}
To demonstrate the effectiveness of our model, we compare to the following recommendation methods.

\textit{Classical methods for implicit feedback}:
\begin{itemize}
\item \textbf{WRMF}, weighted regularized matrix factorization \cite{DBLP:conf/icdm/HuKV08}, which minimizes the square error loss by assigning user rated and unrated items with different confidential values.
\item \textbf{CDAE}, collaborative denoising autoencoder \cite{DBLP:conf/wsdm/WuDZE16}, which utilizes the denoising autoencoder to learn the user hidden representation from implicit feedback.
\end{itemize}

\textit{Methods learning from bag-of-words}:
\begin{itemize}
\item \textbf{CDL}, collaborative deep learning \cite{DBLP:conf/kdd/WangWY15}, is a probabilistic feedforward model for joint learning of stacked denoising autoencoder (SDAE) and collaborative filtering.
\item \textbf{CVAE}, collaborative variational autoencoder \cite{DBLP:conf/kdd/LiS17}, is a generative latent variable model that jointly models the generation of content and rating and uses variational Bayes with inference network for variational inference.
\item \textbf{CML+F}, collaborative metric learning with item features \cite{DBLP:conf/www/HsiehYCLBE17}, which learns a metric space to encode not only users' preferences but also the user-user and item-item similarities.
\end{itemize}

\textit{Methods learning from word sequences}:
\begin{itemize}
\item \textbf{ConvMF}, convolutional matrix factorization \cite{DBLP:conf/recsys/KimPOLY16}, which applies the convolutional neural network (CNN) to capture contextual information of documents and integrates CNN into the probabilistic matrix factorization (PMF).
\item \textbf{JRL}, joint representation learning \cite{DBLP:conf/cikm/ZhangACC17}, is a framework that learns joint representations from different information sources for top-N recommendation.
\end{itemize}

\textit{The proposed method}:
\begin{itemize}
\item \textbf{GATE}, the proposed model, fuses hidden representations from items' ratings and contents by a gating layer, moreover, the word-attention and neighbor-attention are adopted for selecting informative words and learning hidden representations of items' neighborhoods, respectively.
\end{itemize}

Given our extensive comparisons against the state-of-the-art methods, we omit comparisons with methods such as HFT \cite{DBLP:conf/recsys/McAuleyL13}, CTR \cite{DBLP:conf/kdd/WangB11}, SVDFeature \cite{DBLP:journals/jmlr/ChenZLCZY12}, and DeepMusic \cite{DBLP:conf/nips/OordDS13} since they have been outperformed by the recently proposed CDL, CVAE, and JRL.

\subsection{Experiment Settings}
In the experiments, the latent dimension of all the models is set to 50. WRMF adopts the same heuristic weighting function with the proposed model. For CDAE, we follow the settings in the original paper. For CDL, we set $ a = 1 $, $ b =0.01 $, and find that when $ \lambda_{u} = 1 $, $ \lambda_{v} = 10 $, $ \lambda_{n} = 100 $, and $ \lambda_{w} = 0.0001 $ can achieve good performance. For CVAE, the parameters $ a = 1 $, $ b =0.01 $ are the same. When $ \lambda_{u} = 0.1 $, $ \lambda_{v} = 10 $, $ \lambda_{r} = 0.01 $, CVAE can achieve good performance. For CML+F, we follow the author's code to set the margin $ m = 2.0 $, $ \lambda_f = 0.1 $, and $ \lambda_c = 1 $, respectively. The item features are learned by a multi-layer perceptron with a 512-dimensional hidden layer and $ 0.3 $ dropout. For ConvMF, we set the CNN configuration the same as the original paper and find it can achieve a good result when $ a = 1 $, $ b =0.01 $, $ \lambda_{u} = 0.1 $, and $ \lambda_{v} = 10 $. For JRL, we follow the original paper setting to set batch size as $ 64 $, the number of negative samples $ t = 5 $, and $ \lambda_1 = 1 $. The network architectures of above methods are also set the same with the original papers.

For GATE, the ratings of an item is a binary rating vector from all users; the content of an item is the word sequence from its description. We set the maximum length of the word sequence to $ 300 $, and the same setting is also adopted in ConvMF and JRL. Hyper-parameters are set by grid search. The network architecture is set to $ [m, 100, 50, 100, m] $ on all datasets. $ \rho $ is  set to $ 5 $ on citeulike-a, $ 20 $ on movielens-20M, $ 15 $ on Amazon-Books, and $ 20 $ on Amazon-CDs, respectively. $ d_{a} $ is set to $ 20 $, where its effect is shown in section \ref{subsec:parameter_sensitivity}. The learning rate and $ \lambda $ are set to $ 0.01 $ and $ 0.001 $, respectively. The activation function is set to $ tanh $. And the batch size is set to $ 1024 $. Our experiments are conducted with PyTorch\footnote{https://pytorch.org/} running on GPU machines of Nvidia GeForce GTX 1080 Ti\footnote{The code is available on Github: https://github.com/allenjack/GATE}.

\subsection{Performance Comparison}
\begin{figure}[t!]
    \centering
    \begin{subfigure}[t]{0.25\textwidth}
        \centering
        \includegraphics[width=\linewidth]{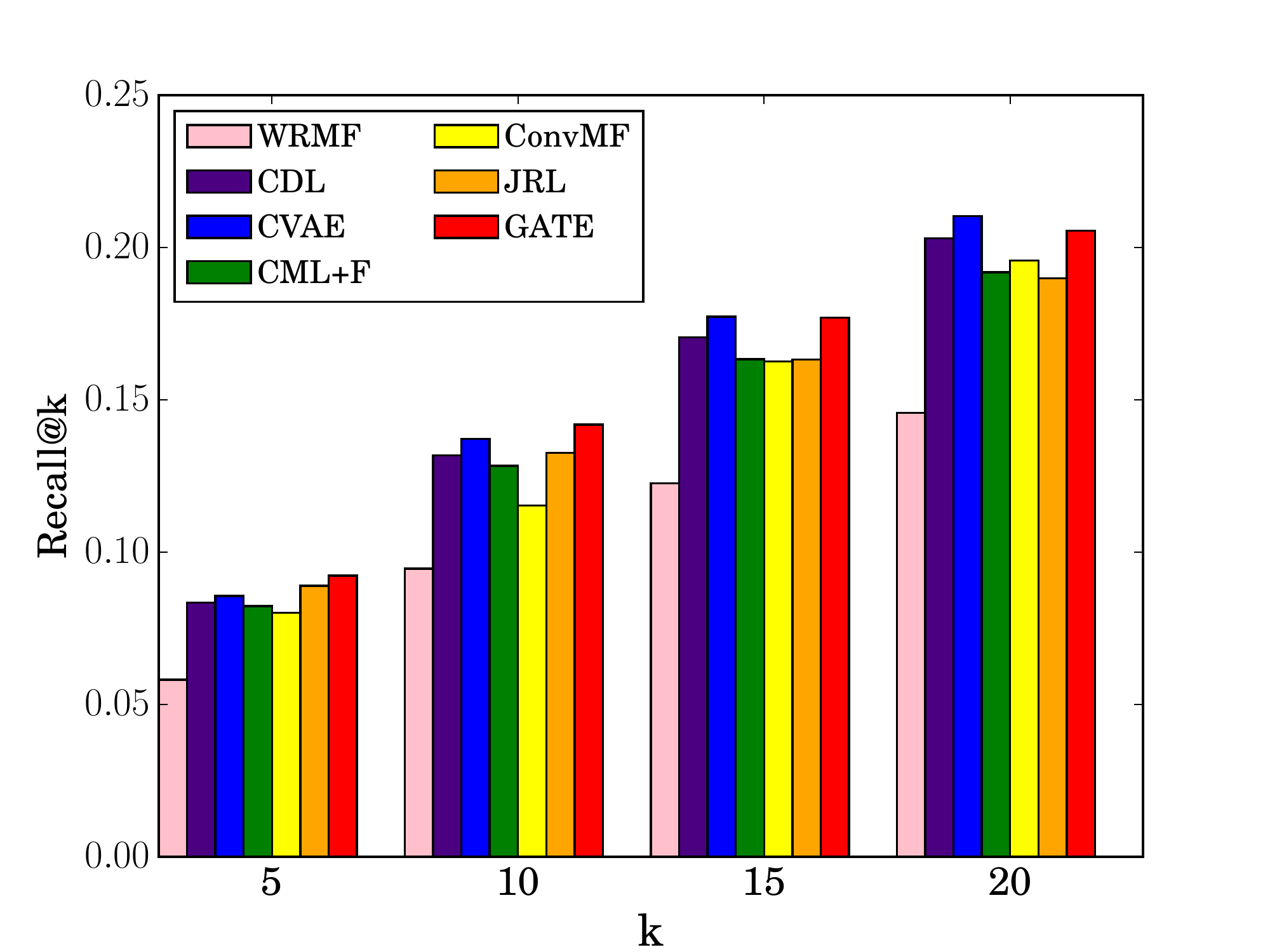}
        \caption{\label{fig:citeulikea_recall} Recall@k on citeulike-a}
    \end{subfigure}%
    \begin{subfigure}[t]{0.25\textwidth}
        \centering
        \includegraphics[width=\linewidth]{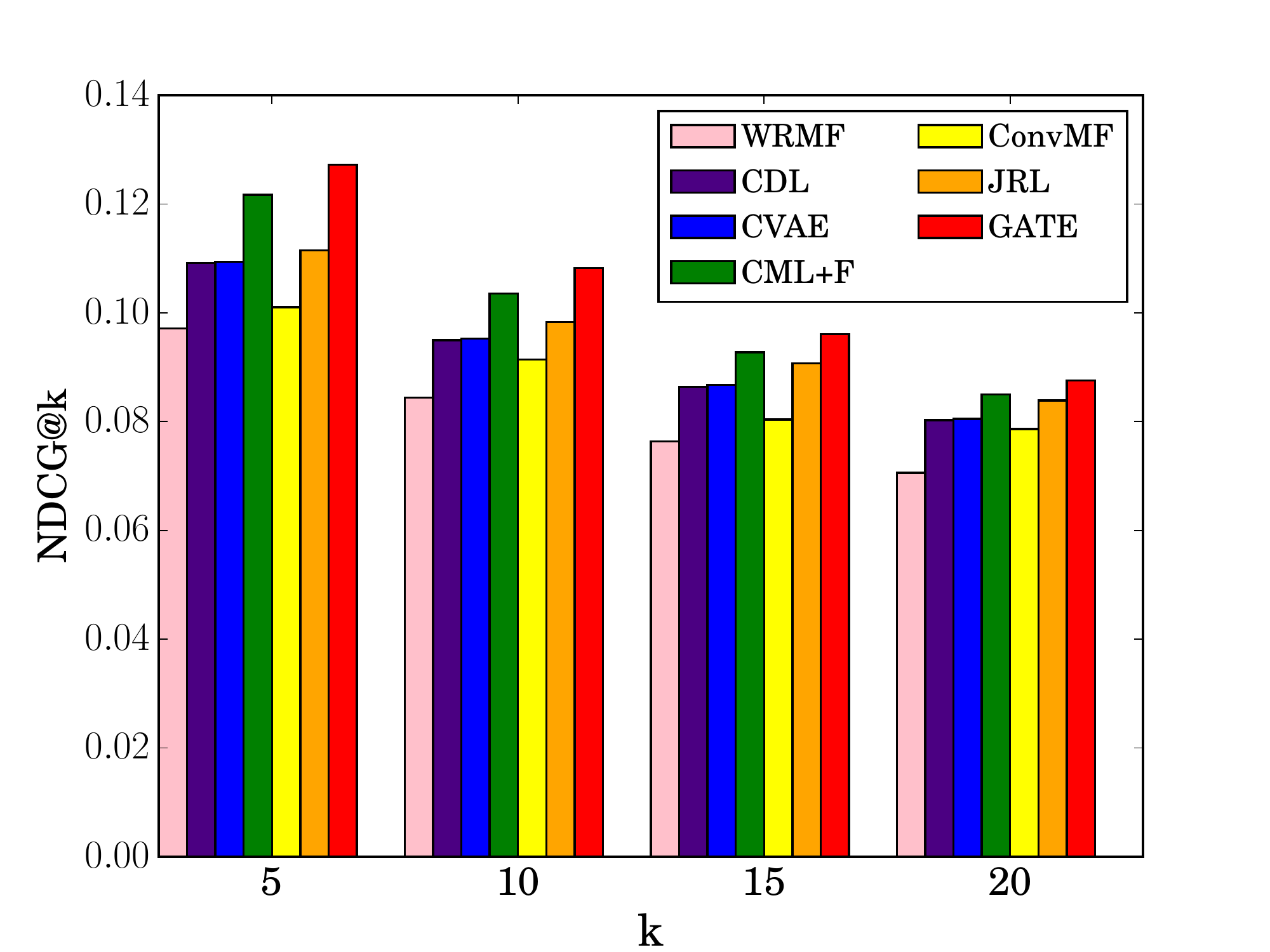}
        \caption{\label{fig:citeulikea_ndcg} NDCG@k on citeulike-a}
    \end{subfigure}
    \caption{\label{fig:citeulikea}The performance comparison on citeulike-a.}
\vspace{-0.5cm}
\end{figure}

\begin{figure}[t!]
    \centering
    \begin{subfigure}[t]{0.25\textwidth}
        \centering
        \includegraphics[width=\linewidth]{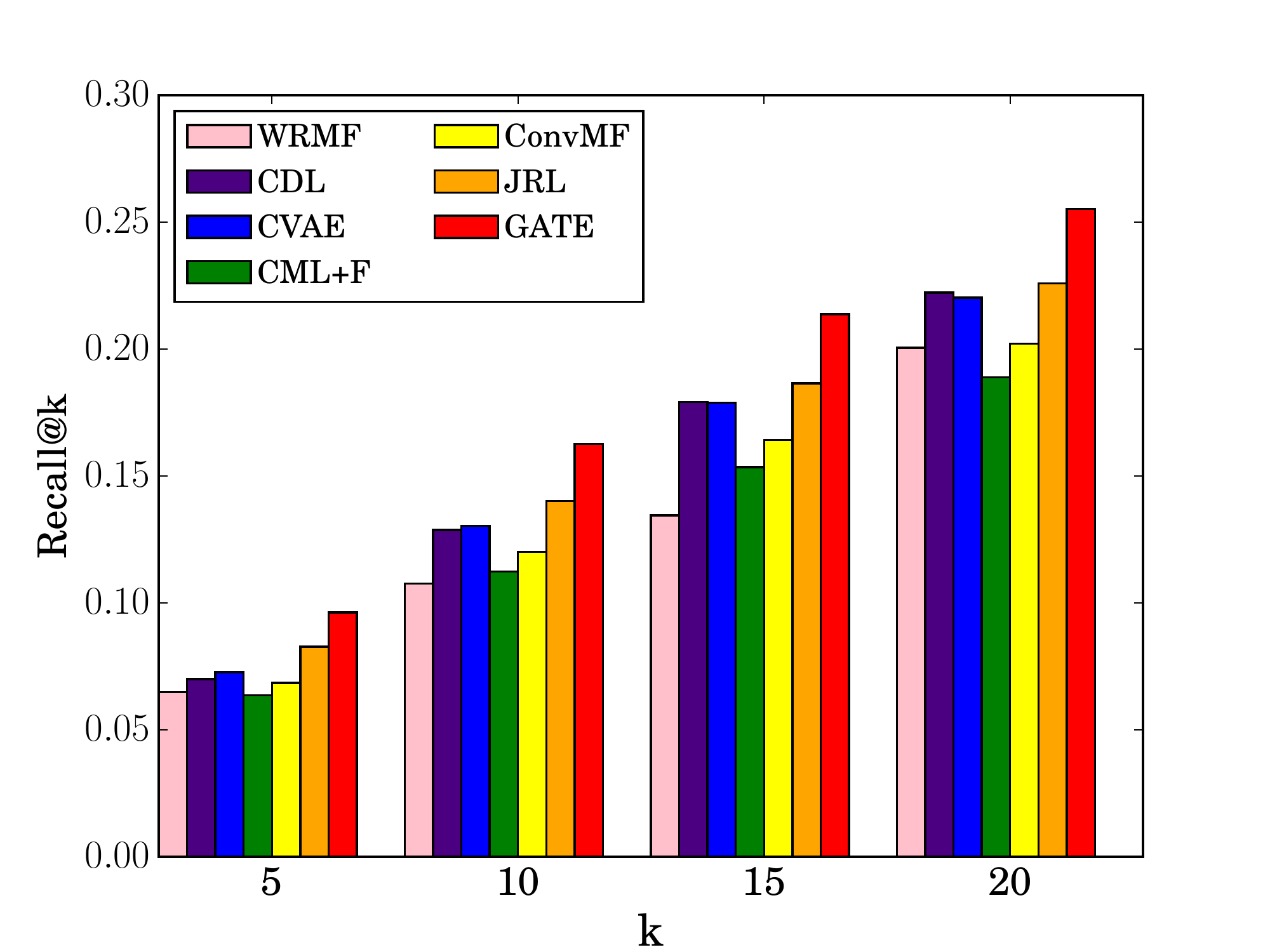}
        \caption{\label{fig:ml20m_recall} Recall@k on movielens-20M}
    \end{subfigure}%
    \begin{subfigure}[t]{0.25\textwidth}
        \centering
        \includegraphics[width=\linewidth]{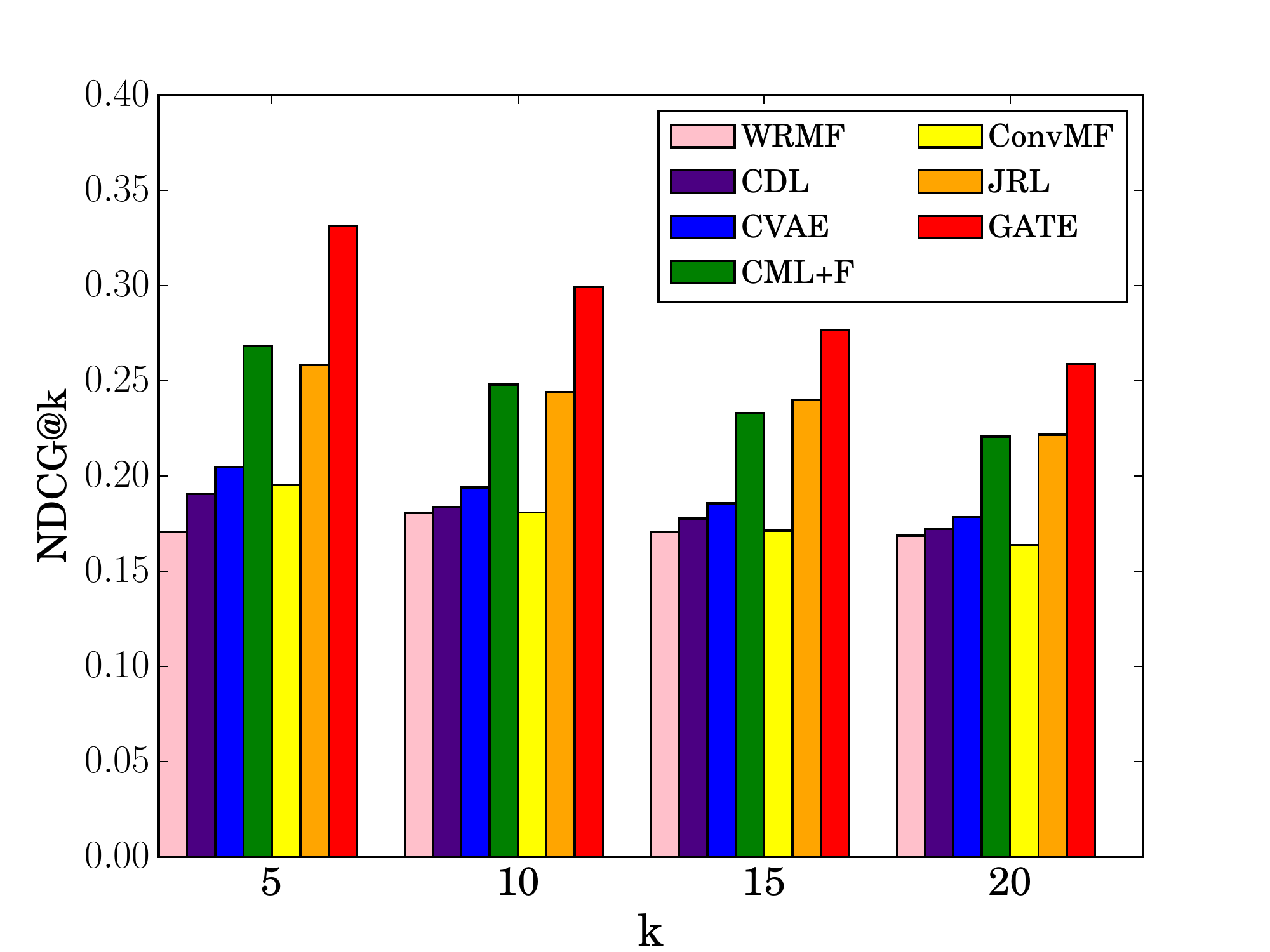}
        \caption{\label{fig:ml20m_ndcg} NDCG@k on movielens-20M}
    \end{subfigure}
    \caption{\label{fig:ml20m}The performance comparison on movielens-20M.}
\vspace{-0.5cm}
\end{figure}

\begin{figure}[t!]
    \centering
    \begin{subfigure}[t]{0.25\textwidth}
        \centering
        \includegraphics[width=\linewidth]{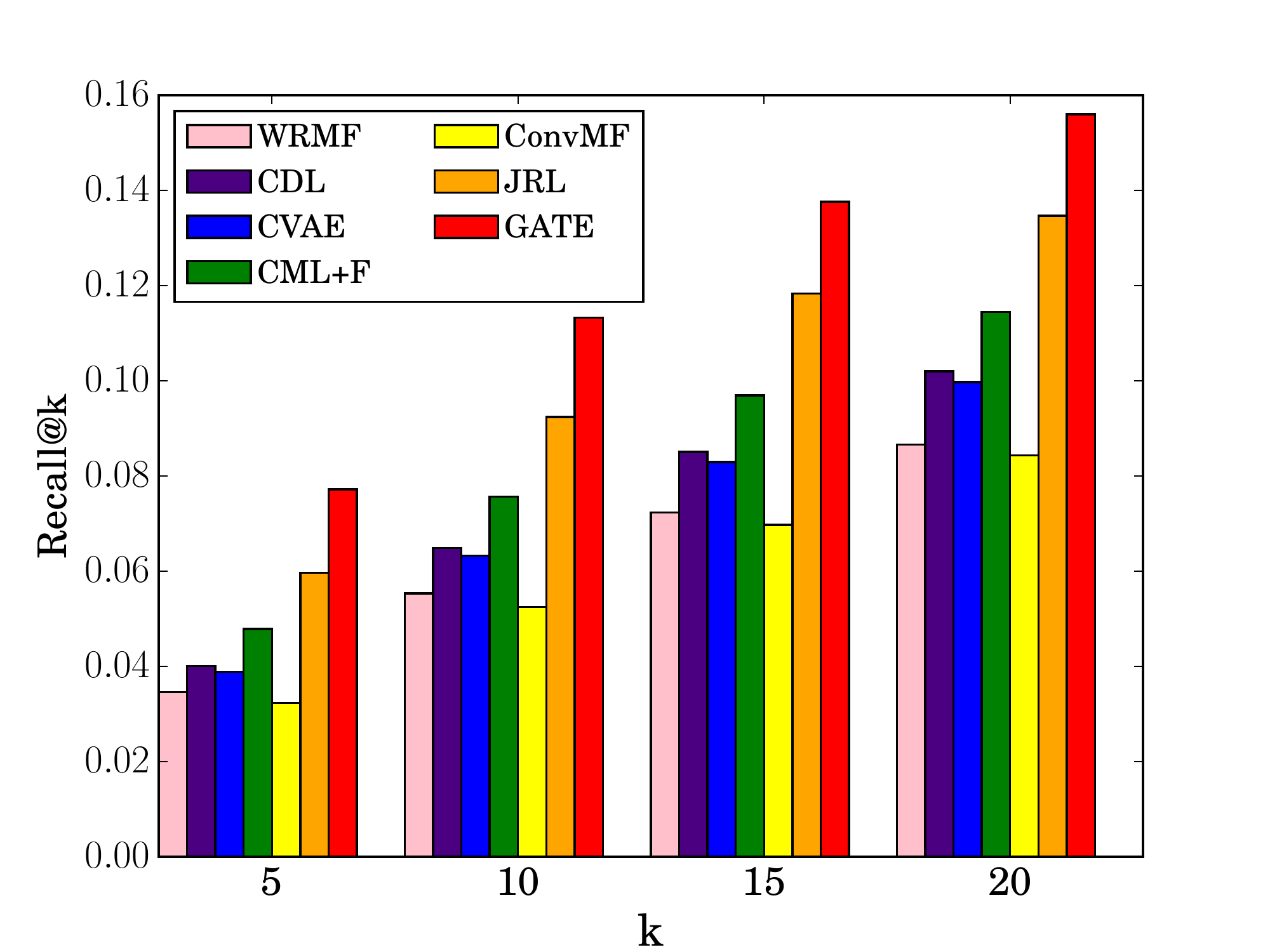}
        \caption{\label{fig:books_recall} Recall@k on Amazon-Books}
    \end{subfigure}%
    \begin{subfigure}[t]{0.25\textwidth}
        \centering
        \includegraphics[width=\linewidth]{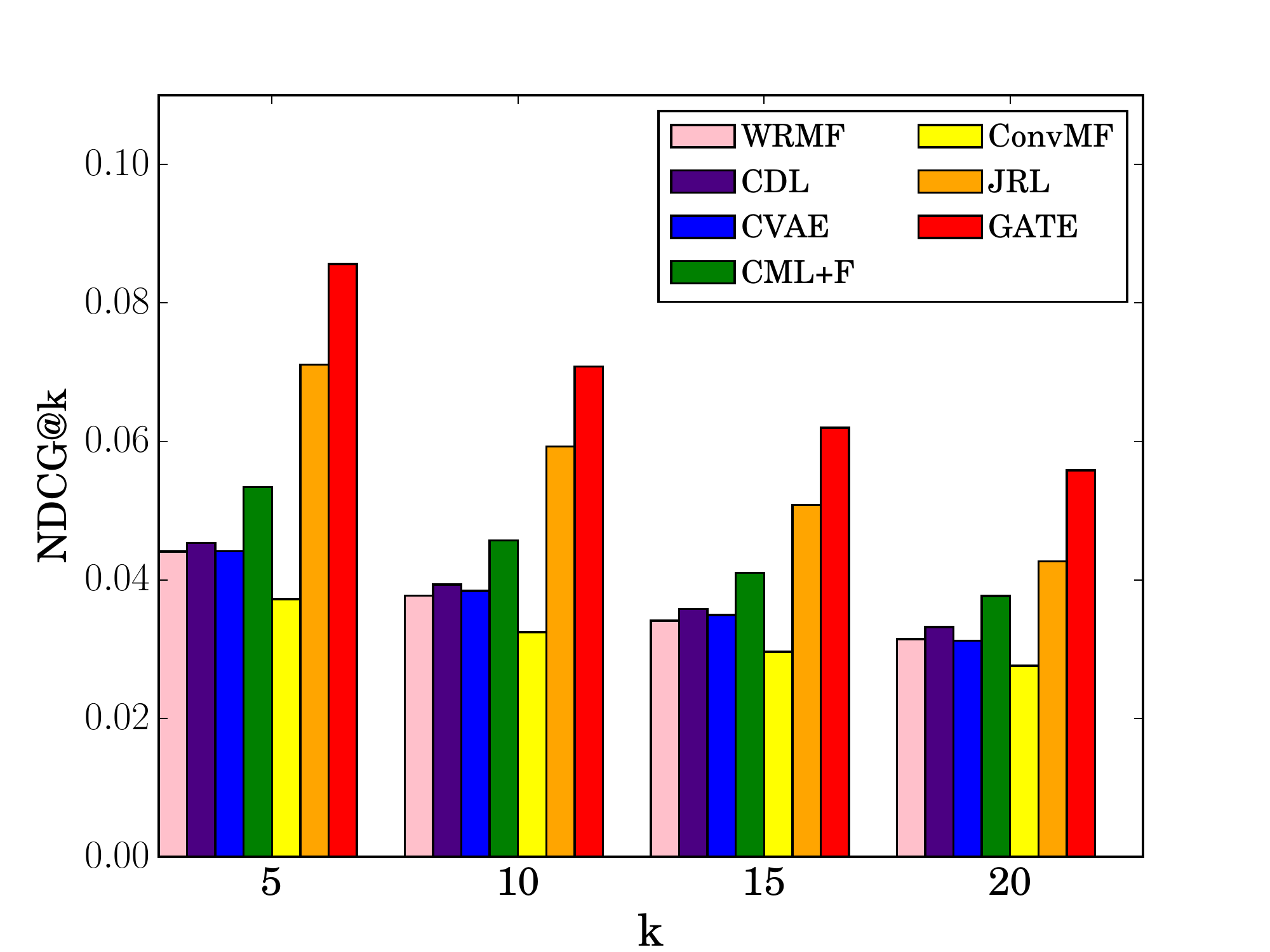}
        \caption{\label{fig:books_ndcg} NDCG@k on Amazon-Books}
    \end{subfigure}
    \caption{\label{fig:books}The performance comparison on Amazon-Books.}
\vspace{-0.5cm}
\end{figure}

\begin{figure}[t!]
    \centering
    \begin{subfigure}[t]{0.25\textwidth}
        \centering
        \includegraphics[width=\linewidth]{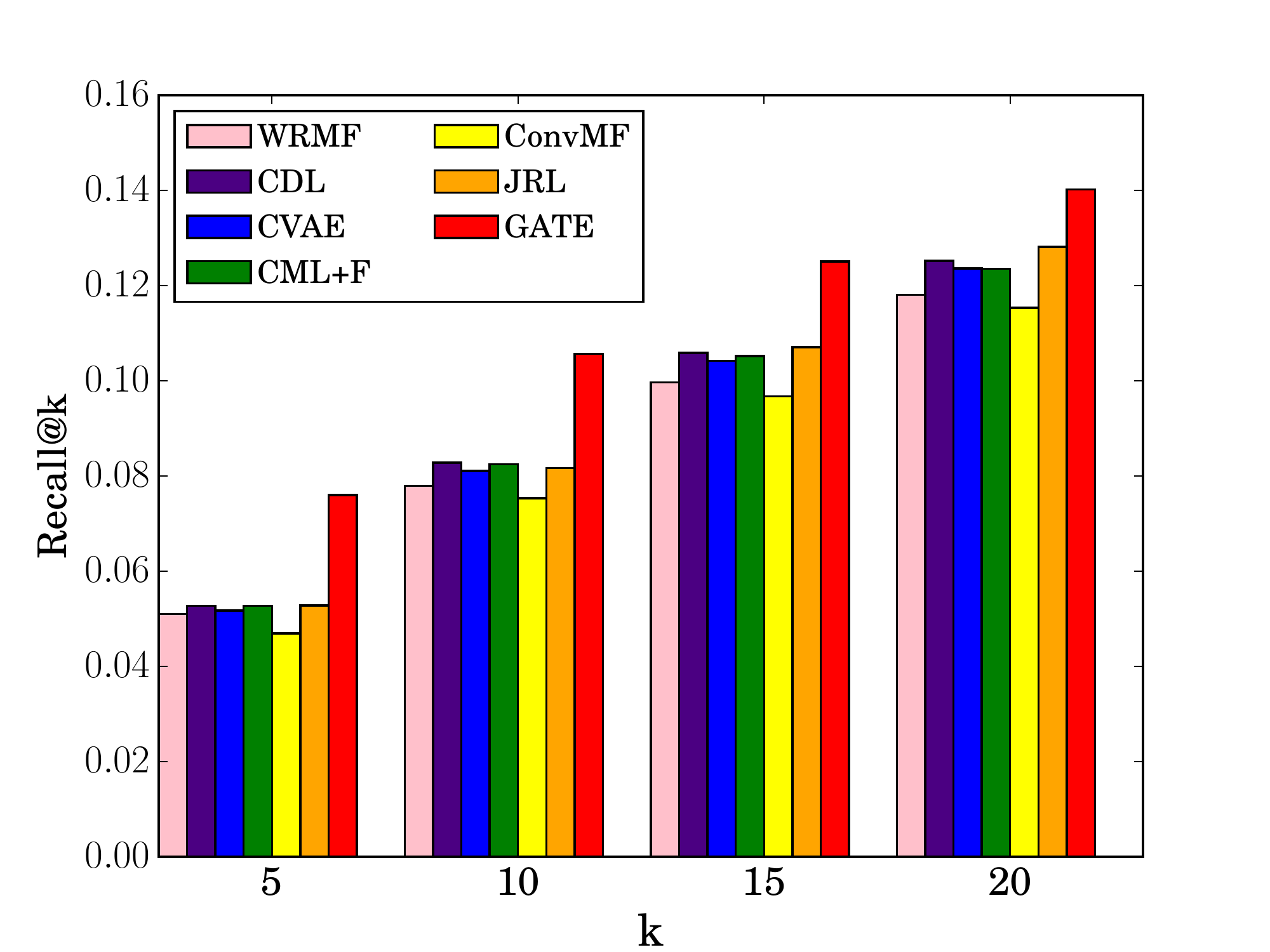}
        \caption{\label{fig:cds_recall} Recall@k on Amazon-CDs}
    \end{subfigure}%
    \begin{subfigure}[t]{0.25\textwidth}
        \centering
        \includegraphics[width=\linewidth]{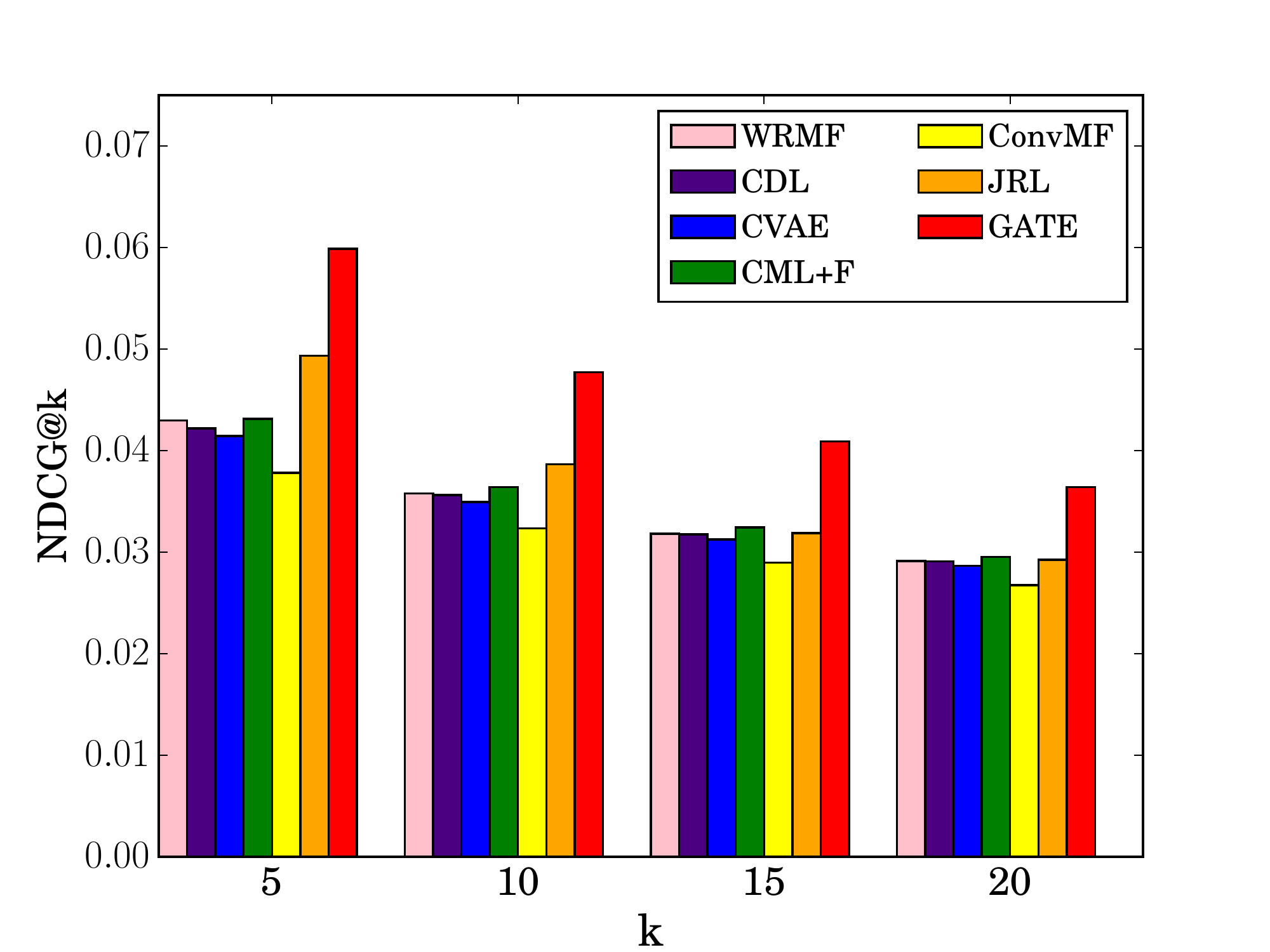}
        \caption{\label{fig:cds_ndcg} NDCG@k on Amazon-CDs}
    \end{subfigure}
    \caption{\label{fig:cds}The performance comparison on Amazon-CDs.}
\vspace{-0.5cm}
\end{figure}

The performance comparison results are shown in Figure \ref{fig:citeulikea}, \ref{fig:ml20m}, \ref{fig:books} and \ref{fig:cds}, and Table \ref{tab:performance_comparison}. Since CDAE is not as good as other state-of-the-art methods when the dataset becomes sparse, we do not present the results of CDAE in the aforementioned figures.

\begin{table*}[ht]
\caption{\label{tab:performance_comparison}The performance comparison of all methods in terms of \textit{Recall@10} and \textit{NDCG@10}. The best performing method is boldfaced. The underlined number is the second best performing method. $ * $, $ ** $, $ *** $ indicate the statistical significance for $ p <= 0.05 $, $ p <= 0.01 $, and $ p <= 0.001 $, respectively, compared to the best baseline method based on the paired t-test. \textit{Improv.} denotes the improvement of our model over the best baseline method.}
\begin{tabular}{|c|c c| c c c| c c| l| c|}
\hline
& \textbf{WRMF} & \textbf{CDAE}  & \textbf{CDL} & \textbf{CVAE} & \textbf{CML+F} & \textbf{ConvMF} & \textbf{JRL} & \textbf{GATE} & \multicolumn{1}{l|}{\textbf{Improv.}} \\\hline
\multicolumn{10}{|c|}{Recall@10} \\
\hline
\textit{citeulike-a} & 0.0946 & 0.0888 & 0.1317 & \underline{0.1371} & 0.1283 & 0.1153 & 0.1325 & \textbf{0.1419} & 3.50\% \\
% \hline
\textit{movielens-20M} & 0.1075 & 0.0751 & 0.1287 & 0.1303 & 0.1123 & 0.1201 & \underline{0.1401} & \textbf{0.1625**} & 15.99\% \\
% \hline
\textit{Amazon-Books} & 0.0553 & 0.0132 & 0.0648 & 0.0632 & 0.0756 & 0.0524 & \underline{0.0924} & \textbf{0.1133*} & 22.62\% \\
% \hline
\textit{Amazon-CDs} & 0.0779 & 0.0191 & \underline{0.0827} & 0.0811 & 0.0824 & 0.0753 & 0.0816 & \textbf{0.1057***} & 27.81\% \\
\hline
\multicolumn{10}{|c|}{NDCG@10} \\ 
\hline
\textit{citeulike-a} & 0.0843 & 0.0736 & 0.0949 & 0.0952 & \underline{0.1035} & 0.0914 & 0.0982 & \textbf{0.1082} & 4.54\% \\
% \hline
\textit{movielens-20M} & 0.1806 & 0.1774 & 0.1836 & 0.1939 & \underline{0.2479} & 0.1807 & 0.2439 & \textbf{0.2992**} & 20.69\% \\ 
% \hline
\textit{Amazon-Books} & 0.0377 & 0.0105 & 0.0393 & 0.0384 & 0.0456 & 0.0324 & \underline{0.0592} & \textbf{0.0708***} & 19.59\% \\
% \hline
\textit{Amazon-CDs} & 0.0357 & 0.0105 & 0.0356 & 0.0349 & 0.0364 & 0.0323 & \underline{0.0386} & \textbf{0.0477***} & 23.58\% \\
\hline
\end{tabular}
\vspace{-0.3cm}
\end{table*}

\textbf{Observations about our model}. First, the proposed model---GATE, achieves the best performance on three datasets with all evaluation metrics, except for the Recall@15 and Recall@20 on citeulike-a, which illustrates the superiority of our model. Second, GATE obtains better results than JRL and ConvMF. Although JRL and ConvMF capture the contextual information in item descriptions by the \textit{doc2vec} model \cite{DBLP:conf/icml/LeM14} and the convolutional neural network, respectively, they equally treat each word of items, which does not consider the effects of informative words, leading to the incomplete understanding of item content information. Third, GATE outperforms CML+F, CVAE, and CDL. The reasons are two-fold: (1) these three methods learn the item content representation through bag-of-words, which neglects the effect that important words can describe the topics or synopses of items; (2) these three methods link the hidden representations from different data sources by a regularization term, which may not smoothly balance the effects of various data representations and incur tedious hyper-parameter tuning. Fourth, GATE achieves better results than WRMF and CDAE, since these two methods do not incorporate the content information, which is crucial when the user-item interaction data is sparse. Fifth, it is important to note that all the compared methods do not consider the user preference on an item's neighborhood, which is captured by the neighbor-attention module of GATE. Sixth, GATE does not significantly improve the performance over other methods on the citeulike-a dataset. One possible reason is that the citeulike-a dataset is relatively small, which makes GATE overfit the data.

\textbf{Other observations}. First, all the results reported on citeulike-a and movielens-20M are better than the results on Amazon-Books and Amazon-CDs, the major reason is that the latter two datasets are more sparse and the data sparsity declines the recommendation performance. Second, JRL and CML+F perform better than other state-of-the-art methods on more sparse datasets. The reason may be that JRL models the contextual information in the item descriptions, which captures the word-word relations in the text. On the other hand, CML+F encodes user-item relationships and user-user/item-item similarities in a joint metric space, which are helpful to find users' preferred items when the data is sparse. Third, although ConvMF models the contextual information from items' descriptions, it still does not perform better than JRL, CML+F, CVAE, and CDL. One possible reason is that the regularization term in ConvMF does not effectively pick up the latent features learned from text to benefit the item latent factors learned from matrix factorization. Fourth, CVAE and CDL achieve similar results on all datasets. One reason is that they have a similar Bayesian probabilistic framework. Fifth, WRMF and CDAE only adopt implicit feedback as input and does not model the auxiliary information, that is why their performance drops when the dataset becomes sparse. In addition, WRMF has similar results with CDL and CVAE on some metrics, which may illustrate that CDL and CVAE may not fully take advantage of the heterogeneous data.

\begin{table}[ht]
\centering
\caption{\label{tab:ablation_analysis}The ablation analysis on Amazon-CDs and Amazon-Books datasets.}
\begin{tabular}{ |l|c|c|c|c| }
\hline
\multirow{2}{*}{Architecture} & \multicolumn{2}{c|}{\textit{CDs}} & \multicolumn{2}{c|}{\textit{Books}} \bigstrut \\\cline{2-5} 
& R@10 & N@10 & R@10 & N@10 \bigstrut \\ 
\hline
(1) stacked AE & 0.0672 & 0.0315 & 0.0745 & 0.0484 \\
(2) reg: AE + W\_Att & 0.0676 & 0.0318 & 0.0304 & 0.0265 \\
(3) gating: AE + W\_Att & 0.0816 & 0.0353 & 0.0793 & 0.0515 \\
(4) gating: AE + GRU & 0.0818 & 0.0352 & 0.0789 & 0.0512 \\
(5) gating: AE + CNN & 0.0777 & 0.0335 & 0.0791 & 0.0495 \\
(6) GATE & \textbf{0.1057} & \textbf{0.0477} & \textbf{0.1133} & \textbf{0.0708} \\
\hline
\end{tabular}
\vspace{-0.3cm}
\end{table}

\subsection{Ablation Analysis}

\begin{table*}[ht]
\centering
\caption{\label{tab:neighbor_attention}A case study of the importance scores computed by the neighbor-attention module. The number inside $ (\cdot) $ indicates the number of \textit{fluctuation}'s occurrences excluding references in an article.}
\begin{tabular}{ |l|p{11cm}|c| }
 \hline
 Target & Neighbor & Score \\
 \hline
 %\multirow{4}{*}{The metabolic world of Escherichia coli is not small} & Community structure in social and biological networks  & 0.06364  \bigstrut \\\cline{2-3}  
 % & Exploring complex networks  & 0.46817  \bigstrut \\\cline{2-3}
 % & Reconstruction of metabolic networks from genome data and analysis of their global structure for various organisms  & \textbf{0.46819}  \bigstrut \\ 
 %\hline
 \multirow{5}{*}{Fluctuations in network dynamics} & Genomic analysis of regulatory network dynamics reveals large topological changes (0) & 0.07172  \bigstrut \\\cline{2-3} 
  & Frequency of occurrence of numbers in the World Wide Web (10) & 0.22090 \bigstrut \\\cline{2-3} 
  & Complex networks: Structure and dynamics (16)  & 0.26835 \bigstrut \\\cline{2-3}
  & Noise in protein expression scales with natural protein abundance (\textbf{36}) & \textbf{0.43903} \bigstrut \\ 
 \hline
\end{tabular}
\vspace{-0.3cm}
\end{table*}

To verify the effectiveness of the proposed word-attention, gating layer, and neighbor-attention modules, we conduct an ablation analysis in Table \ref{tab:ablation_analysis} to demonstrate the performance each module contributes to the GATE model. In (1), we utilize the weighted stacked AE without any other components. In (2), we regularize $ \mathbf{z}_i^r $ and $ \mathbf{z}_i^c $ by $ L2 $ norm on the top of (1), following the same manner in \cite{DBLP:conf/kdd/WangWY15,DBLP:conf/kdd/LiS17}. We tried the regularization parameters $ \{ 0.01, 0.1, 0.5, 1, 10 \} $, where $ 0.1 $ gives the best results. In (3), we plug the gating layer to connect $ \mathbf{z}_i^r $ and $ \mathbf{z}_i^c $ on the top of (1). In (4), we adopt the a recurrent neural network structure--gated recurrent units (GRUs) \cite{DBLP:conf/emnlp/ChoMGBBSB14} to learn $ \mathbf{z}_i^c $, which is also linked to $ \mathbf{z}_i^r $ by the proposed gating layer. In (5), we replace the GRUs in (4) with a convolutional neural network (CNN), where the structure and hyper-parameters are set the same in \cite{DBLP:conf/recsys/KimPOLY16}. In (6), we present the overall GATE model to show the significance of the neighbor-attention module.

From the results shown in Table \ref{tab:ablation_analysis}, we have some observations. First, from (2) and (3), the gating layer achieves better results than regularization. One possible reason is that the neural gate can extract representative parts and mask off insignificant parts from the input hidden representations. Second, from (3), (4), and (5), we observe that our word-attention module has similar performance with GRUs and CNNs but with \textit{fewer} parameters\footnote{We verified the number of parameters of all three models by the \textit{named\_parameters()} function provided by PyTorch.} (if we set the word embedding size to 50 ($ h = 50 $), then the number of learned parameters of our word-attention module is 3,590, the number of parameters of the one-recurrent-layer GRU is 15,300, the number of parameters of the CNN in \cite{DBLP:conf/recsys/KimPOLY16} is 75,350). This result demonstrates that the proposed word-attention module can effectively learn the item hidden representation from items' descriptions. Third, from (1), (3), and (6), we observe that our neighbor-attention may play a critical role in the overall model. The results demonstrate that modeling users' preferences on an item's neighborhood is an effective supplementary for inferring their preferences on this item.

%If we set the word embedding size to 50 ($ h = 50 $), then the number of learned parameters of our word-attention module is 50 * 50 + 50 + 50 * 20 + 20 + 20 = 3590, the number of parameters of GRU is 150 * 50 + 150 * 50 + 150 + 150 = 15300, the number of parameters of CNN is 

\subsection{The Sensitivity of Hyper-parameters} \label{subsec:parameter_sensitivity}
\begin{figure}[t!]
    \centering
    \begin{subfigure}[t]{0.25\textwidth}
        \centering
        \includegraphics[width=\linewidth]{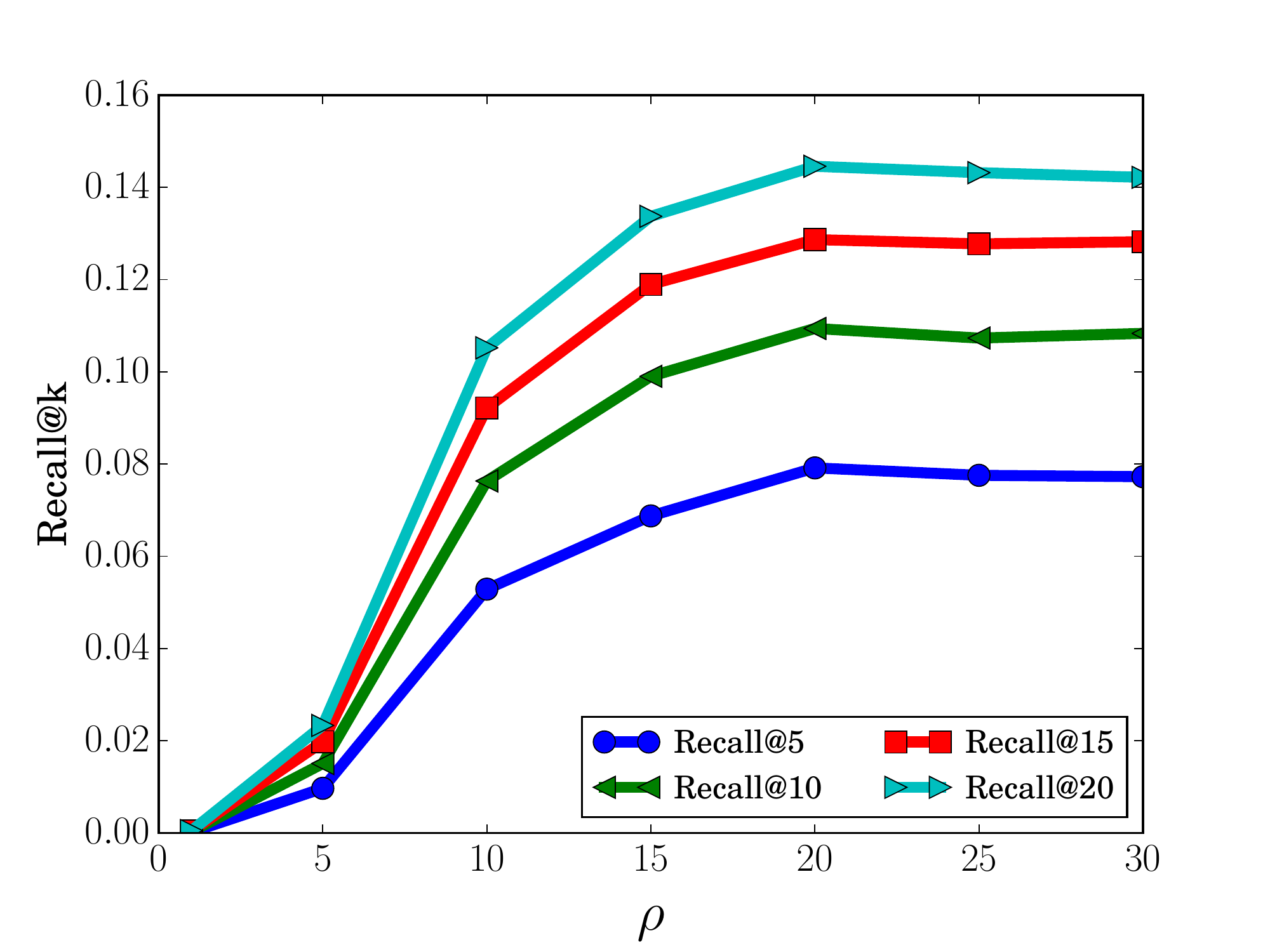}
        \caption{\label{fig:rating_weight} $ \rho $ on Amazon-CDs}
    \end{subfigure}%
    \begin{subfigure}[t]{0.25\textwidth}
        \centering
        \includegraphics[width=\linewidth]{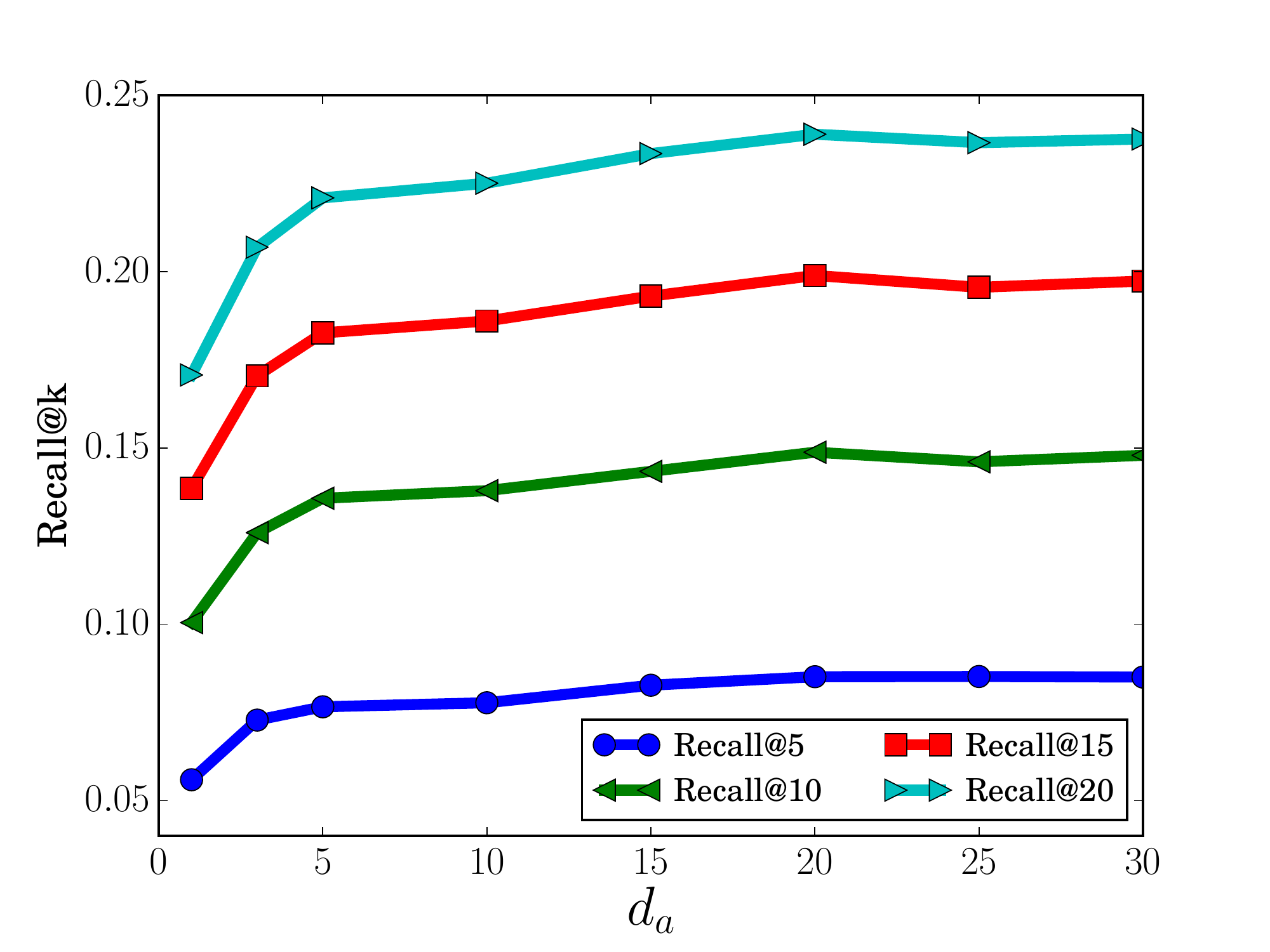}
        \caption{\label{fig:da_var}$ d_{a} $ on movielens-20M}
    \end{subfigure}
    \caption{\label{fig:hyper_parameter}The effects of $ \rho $ and $ d_{a} $.}
\vspace{-0.5cm}
\end{figure}

The effects of $ \rho $ and $ d_{a} $ are shown in Figure \ref{fig:hyper_parameter}, which have similar trends on other datasets. We can observe that with the increase of $ \rho $, the performance improves and becomes stable. The reason is that the larger value of $ \rho $ makes the model concentrate more on the items that users interacted with before, where users' preferences are more accurately captured. For the variation of $ d_{a} $, we verify that utilizing a vector to measure the importance of a word is more effective than just using a single value in our scenario because the score vector describes the relations between each word from different aspects. Note that we do not include the neighbor-attention module when testing the effect of $ d_a $.

\begin{table}[ht]
\centering
\caption{\label{tab:word_attention_case}A case study of the word-attention.}
\begin{tabular}{|p{8.5cm}|}
 \hline
 The Summary of Article 16797 in \textit{citeulike-a} \\
 \hline
  We \colorbox[cmyk]{0.22899408,0,0,0}{present} the \colorbox[cmyk]{0.05,0,0,0}{first} \colorbox[cmyk]{0.47920801,0,0,0}{parallel} \colorbox[cmyk]{0.42350478,0,0,0}{implementation} of the T-Coffee consistency-based \colorbox[cmyk]{0.44287665,0,0,0}{multiple} \colorbox[cmyk]{0.91484752,0,0,0}{aligner}. We \colorbox[cmyk]{0.47988166,0,0,0}{benchmark} it on the \colorbox[cmyk]{0.60690942,0,0,0}{Amazon} \colorbox[cmyk]{0.5767137,0,0,0}{Elastic} \colorbox[cmyk]{1,0,0,0}{Cloud} (EC2) and \colorbox[cmyk]{0.07614019,0,0,0}{show} that the \colorbox[cmyk]{0.47920801,0,0,0}{parallelization} \colorbox[cmyk]{0.64223942,0,0,0}{procedure} is \colorbox[cmyk]{0.66585344,0,0,0}{reasonably} \colorbox[cmyk]{0.14790168,0,0,0}{effective}. We also \colorbox[cmyk]{0.17824306,0,0,0}{conclude} that for a \colorbox[cmyk]{0.73873464,0,0,0}{web} \colorbox[cmyk]{0.62376878,0,0,0}{server} with \colorbox[cmyk]{0.54477924,0,0,0}{moderate} \colorbox[cmyk]{0.45925353,0,0,0}{usage} (10K hits/month) the \colorbox[cmyk]{1,0,0,0}{cloud} \colorbox[cmyk]{0.1526081,0,0,0}{provides} a \colorbox[cmyk]{0.41543013,0,0,0}{cost-effective} \colorbox[cmyk]{0.11651343,0,0,0}{alternative} to in-house deployment. \\
 \hline
\end{tabular}
\begin{tabular}{|p{8.5cm}|}
 \hline
 The Summary of Article 120 in \textit{citeulike-a} \\
 \hline
  We \colorbox[cmyk]{0.26519101,0,0,0}{identify} a \colorbox[cmyk]{0.97783526,0,0,0}{metaphor} for the \colorbox[cmyk]{0.14705531,0,0,0}{design} \colorbox[cmyk]{0.05,0,0,0}{activity}: we \colorbox[cmyk]{0.27167728,0,0,0}{view} \colorbox[cmyk]{0.14705531,0,0,0}{design} as bricolage. We \colorbox[cmyk]{0.56098289,0,0,0}{start} from \colorbox[cmyk]{0.66025667,0,0,0}{describing} bricolage, and we \colorbox[cmyk]{0.55769996,0,0,0}{proceed} to the \colorbox[cmyk]{0.29891564,0,0,0}{relationship} of \colorbox[cmyk]{0.14705531,0,0,0}{design} to \colorbox[cmyk]{0.65675487,0,0,0}{art}. We \colorbox[cmyk]{0.42574612,0,0,0}{obtain} a characterisation of \colorbox[cmyk]{0.14705531,0,0,0}{design} that \colorbox[cmyk]{0.45614803,0,0,0}{enables} us to \colorbox[cmyk]{0.50011938,0,0,0}{show} that both \colorbox[cmyk]{0.33460008,0,0,0}{traditional} and \colorbox[cmyk]{0.44535416,0,0,0}{contemporary} \colorbox[cmyk]{0.14705531,0,0,0}{design} are \colorbox[cmyk]{0.24131516,0,0,0}{forms} of bricolage. We \colorbox[cmyk]{0.63579387,0,0,0}{examine} the \colorbox[cmyk]{0.74485674,0,0,0}{consequences} of '\colorbox[cmyk]{0.14705531,0,0,0}{design} as bricolage' for the \colorbox[cmyk]{0.29891564,0,0,0}{relationship} between \colorbox[cmyk]{0.14705531,0,0,0}{design} and \colorbox[cmyk]{0.83735575,0,0,0}{science} and for the \colorbox[cmyk]{0.48437127,0,0,0}{extent} of the \colorbox[cmyk]{0.14705531,0,0,0}{design} \colorbox[cmyk]{0.05,0,0,0}{activity}. \\
 \hline
\end{tabular}
\vspace{-0.3cm}
\end{table}

\subsection{Word- and Neighbor-Attention Case Studies}

To visualize the word-attention effects, we sum along the first dimension of $ \mathbf{A}_{i} \in \mathbb{R}^{d_a \times l_i} $ (Eq. \ref{eq:attention_matrix}) to get $ \mathbf{a}_{i} \in \mathbb{R}^{l_i} $, which can be treated as the accumulated attention weights of each word. For the ease of visualization, we normalize $ \mathbf{a}_{i} $ following the same procedure in \cite{DBLP:journals/corr/LinFSYXZB17} and words with lower scores are not colored. Two examples of word-attention visualization are shown in Table \ref{tab:word_attention_case}. From the first example, we can observe that the words \textit{aligner} and \textit{cloud} have the highest importance scores, which may reflect the topic and platform of this paper. On the other hand, the words \textit{present}, \textit{show}, and \textit{conclude} are widely used in all the papers, which are less attractive. In the second example, the situation is the same. The most important word that selected by the word-attention is \textit{metaphor}, which may reveal the motif of the article. % The word \textit{bricolage} does not exist in the dataset vocabulary, that is why our model does not highlight it.

The neighbor-attention case study is shown in Table \ref{tab:neighbor_attention}. The neighbors of the target article are provided by the citation graph of the citeulike-a dataset. From this case, we observe that the neighbor attention score can identify an item's important neighbors. In the example, the target article finds a scaling rule in network dynamics, and the fourth neighbor of the target also observes the same scaling behavior for all groups of genes. If we treat \textit{fluctuation} as the key topic of the target article, the number of \textit{fluctuation}'s occurrences in the target's neighbors may reveal how related the target with its neighbors. We also list the count of \textit{fluctuation} after the article title. The counts of \textit{fluctuation} further verify the importance scores computed by our neighbor-attention module.

\section{Conclusion}
In this paper, we proposed a gated autoencoder with the word- and neighbor-attention. The model learned items' hidden representations from ratings and contents in a gated manner. Moreover, the model also captured items' informative words and representative neighbors by word- and neighbor-attention modules, respectively. Experimental results on four real-world datasets clearly validated the performance of our model over many state-of-the-art methods and showed the effectiveness of the gating and attention modules.

\bibliographystyle{ACM-Reference-Format}
\bibliography{gat} 

\end{document}